\documentclass[reprint,amsmath,amssymb,aps]{revtex4-2}
\usepackage{graphicx}
\usepackage{dcolumn}
\usepackage{bm}
\usepackage{amsmath}
\usepackage{amssymb}
\usepackage{txfonts}
\usepackage{mathrsfs}
\usepackage{xcolor}
\usepackage[mathlines]{lineno}

\usepackage{gensymb}
\usepackage{float}

\def\b0{{\mathbf 0}}

\def\b0{{\mathbf 0}}

\def\beq{\begin{equation}}
\def\eeq{\end{equation}}

\begin{document}

\title{Phase diagram of two-component mean-field Bose mixtures}
\author{Oskar Stachowiak  }
\email{oskar.stachowiak@fuw.edu.pl}
\affiliation{Institute  of Theoretical Physics, Faculty of Physics, University of Warsaw, Pasteura 5, 02-093 Warsaw, Poland}
\author{Pawel Jakubczyk } 
\email{pawel.jakubczyk@fuw.edu.pl}
\affiliation{Institute  of Theoretical Physics, Faculty of Physics, University of Warsaw, Pasteura 5, 02-093 Warsaw, Poland}

\date{\today}

\begin{abstract} 
 We revisit the structure of the phase diagram of the two-component mean-field Bose mixture at finite temperatures, considering both the cases of attractive and repulsive interspecies interactions. In particular, we analyze the evolution of the phase diagram upon driving the system towards collapse and point out its distinctive features in this limit. We provide analytical insights into the global structure of the phase diagram and the properties of the phase transitions between the normal phase and the phases involving Bose-Einstein condensates. \emph{Inter alia} we analytically demonstrate that for sufficiently weak interspecies interactions $a_{12}$ the system generically exhibits a line of quadruple points but has no triple nor tricritical points in the phase diagram spanned by the chemical potentials $\mu_1$, $\mu_2$ and temperature $T$. In contrast, for sufficiently large, positive values of $a_{12}$, the system displays both triple and tricritical points but no quadruple points. As pointed out in recent studies, in addition to the phase transitions involving condensation, the mixture may be driven through a liquid-gas type transition, and we clarify the conditions for its occurrence. We finally discuss the impact of interaction- and mass-imbalance on the phase diagram of the mixture.

\end{abstract}

\maketitle

\section{Introduction} 
Bose mixtures received an explosion of interest in the years following the realization of Bose-Einstein condensation (BEC) in trapped ultracold atomic gases, and their intense exploration continues up to now \cite{Ho_1996, Esry_1997, Ohberg_1998, Timmermans_1998, Pu_1998, Ao_1998, Hall_1998, Shi_2000, Riboli_2002, Altman_2003, Mertes_2007, Bhongale_2008, Papp_2008, Catani_2008, Anderson_2009, Gadway_2010, Hubener_2009, Capograsso_2010, McCarron_2011, Facchi_2011, Schaeybroeck_2013, Wang_2013, Lv_2014, Ceccarelli_2015, Lingua_2015, Polo_2015, Roy_2015, Ceccarelli_2016, Wang_2016, Abulseoud_2016, Landig_2016, Lee_2018, Utesov_2018, Bisset_2018, Boudjemaa_2018, Boudjemaa_2020, Ota_2019, Ota_2020, He_2020, Pyzh_2020, Isaule_2021, Lee_2021, Serrano_2021, Isaule_2022, Rakhimov_2022, Spada_2023, Spada_2023_2, Liu_2023, Jakubczyk_2024, Jakubczyk_2024_2, Chen_2024}. Despite this, a number of key features concerning the phase diagram of such systems have been unveiled only very recently. For the most elementary case of interaction- and mass-balanced mixtures with simple repulsive interactions, these include the first-order character of the transition to the BEC states occurring in substantial portions of the phase diagram, and previously unexplored liquid-gas type phase transitions within the regions of the phase diagram occupied by the non-condensed phases. It seems plausible that density jumps associated with such transitions may be detectable in current cold-atom experiments, in particular considering that their exploration does not require cooling the system below the condensation temperature. Apart from the above-mentioned aspects of the phase diagram, which were neglected or overlooked in the early theoretical studies, there are other properties of Bose mixtures, which, up to date, remain hardly explored theoretically and are also likely to become of growing interest and relevance in near future experimental investigations. These concern in particular the case of negative (attractive) interspecies coupling and the limit, where the system is driven towards collapse \cite{Kagan_1998, Sackett_1998, Cornish_2000, Abdullaev_2003, Mardonov_2015}  by tuning interactions. Another aspect of Bose mixtures that has not been explored much relates to the effect of interaction and mass imbalance on the phase diagram structure as well as thermodynamic and excitation spectra properties.
 
 In the present paper, we aim to clarify these points systematically within an exactly soluble model describing a gas of bosonic particles interacting via a mean-field type potential. We build upon the recent study of Ref.~\cite{Jakubczyk_2024}, where the presently implemented model was introduced and employed to demonstrate, largely by numerical means, the possibility of realizing Bose-Einstein condensation as a first-order transition; explore the related multicritical behaviours; as well as detect the new liquid-gas type transition between non-condensed states of the system. The present work supplements and extends that previous contribution both by delivering systematic analytical insights and by exploring relevant parameter ranges not analyzed previously, the case of attractive interactions and strongly imbalanced setups in particular.
 



Although many aspects of the phase diagram of homogeneous two-component Bose-Bose mixtures were studied before \cite{Ho_1996, Esry_1997, Ohberg_1998, Timmermans_1998, Pu_1998, Ao_1998, Hall_1998, Shi_2000, Riboli_2002, Altman_2003, Mertes_2007, Bhongale_2008, Papp_2008, Catani_2008, Anderson_2009, Gadway_2010, Hubener_2009, Capograsso_2010, McCarron_2011, Facchi_2011, Schaeybroeck_2013, Wang_2013, Lv_2014, Ceccarelli_2015, Lingua_2015, Polo_2015, Roy_2015, Ceccarelli_2016, Wang_2016, Abulseoud_2016, Landig_2016, Lee_2018, Utesov_2018, Bisset_2018, Boudjemaa_2018, Boudjemaa_2020, Ota_2019, Ota_2020, He_2020, Pyzh_2020, Isaule_2021, Lee_2021, Serrano_2021, Isaule_2022, Rakhimov_2022, Spada_2023, Spada_2023_2, Liu_2023, Jakubczyk_2024, Jakubczyk_2024_2, Chen_2024}, the general picture emergent from available literature is scattered, largely incomplete, and occasionally misleading. In addition, our work offers general analytical insights, whereas the majority of earlier analyses relied on numerical results restricted to specific parameter sets.

 The paper is organized as follows: 
 In Sec.~II, we introduce the considered model and review the previously obtained results. We discuss the condition of the system stability and its limitations if the interspecies interaction $a_{12}$ becomes attractive ($a_{12}<0$). In Sec.~III, we analyze the structure of the phase diagram for $a_{12}<0$. We in particular resolve the limit $a_{12}\to {{a_{12}^{(0)}}}^+$, where the system approaches instability. It turns out that this limit is reflected in specific geometric features of the phase diagram. In Sec.~IV, we analyze the conditions for realizing condensation as a first-order phase transition and elaborate on the associated question of the presence/absence of tricritical points on the phase diagram. We provide exact (at mean-field level) statements concerning this problem. Sec. V deals with the liquid-gas-type phase transition occurring within the portion of the phase diagram which does not involve condensates. We formulate mean-field exact criteria for the occurrence of this phase transition; in particular, we demonstrate that they always appear for high temperatures and/or interspecies interactions. Sec.~VI is concerned with the global properties of the phase diagram and the number of triple points that may exist in  fixed-$T$ projections of the phase diagram. We, in particular, point out the disagreement between earlier proposed phase diagram structures and our general findings. In Sec.~VII, we discuss the new features of the phase diagram that appear due to interaction and/or mass imbalance. Sec. VIII contains a summary and outlook.
\section{Model and review of previous results} 
We consider the model of mean-field interacting spinless bosons introduced in \cite{Jakubczyk_2024} and defined by the Hamiltonian
\begin{equation} 
\label{imp_B_mix}
\hat{H}=\sum_{\vec{k},i}\epsilon_{\vec{k},i}\hat{n}_{\vec{k},i}+\sum_{i, j}\frac{a_{i,j}}{2V}\hat{N}_i \hat{N}_{j}\;.
\end{equation}
Here $i,j\in\{1,2\}$, the intraspecies couplings 
$a_{i,i}>0$ (hereafter denoted as $a_i$) are assumed positive, while the interspecies coupling  $a_{1,2}=a_{2,1}$ (denoted hereafter as $a_{12}$) has no restriction on its sign. We will in particular, in Sec.~III, analyze the instability occurring when $a_{12}$ is negative and becomes sufficiently large in absolute value. The single-particle states for each type of particles are labelled by the wavevector $\vec{k}$, $\hat{n}_{\vec{k},i}$ are the corresponding occupation number operators, $\hat{N}_i:=\sum_{\vec{k}}\hat{n}_{\vec{k},i}$ denote the total particle number operators for each of the species, and the dispersion takes the standard form $\epsilon_{\vec{k},i}=\hbar^2\vec{k}^2/(2m_i)$. The system is subject to periodic boundary conditions, maintained at fixed temperature $k_B T=\beta^{-1}$ and enclosed in a three-dimensional cubic vessel of volume $V$. We will use the grand canonical ensemble, where the chemical potentials $\mu_1$, $\mu_2$ of the mixture constituents are fixed, and the grand-canonical free energy density $\omega(T,\mu_1, \mu_2)$ follows from the grand canonical partition function $\Xi(T,V,\mu_1,\mu_2)$ via 
\begin{equation}
\omega(T,\mu_1, \mu_2)=-\lim_{V\to\infty}\frac{1}{V\beta}\ln\Xi(T,V,\mu_1,\mu_2)\;.    
\end{equation}  
The construction of the model represents a rather straightforward extension of the so-called imperfect Bose gas model \cite{Davies_1972, Buffet_1983, Berg_1984, Lewis_1985, Smedt_1986, Zagrebnov_2001, Napiorkowski_2011, Napiorkowski_2013, Jakubczyk_2013_2, Jakubczyk_2016, Diehl_2017,  Jakubczyk_2018, Lebek_2020, Frerot_2022} to the case of binary mixtures. We note in particular that the mean-field interaction term $\sim N^2/V$  may be derived from a realistic two-body interaction potential via the so-called Kac scaling procedure \cite{Hemmer_1976}, where the interaction is progressively scaled to be extremely weak and long-ranged at the same time.  
We also point out \cite{Jakubczyk_2024} that the structure of the self-consistency (saddle-point)  equations deriving from the model of Eq.~(\ref{imp_B_mix}) is, up to redefinition of parameters, equivalent to the one arising from the Hartree-Fock treatment \cite{Schaeybroeck_2013} of the weakly interacting Bose gas in the dilute limit. We therefore expect that the conclusions of the present study are not restricted to the setup involving extremely long-ranged interactions and may be pertinent to realistic situations. 

As demonstrated in Ref.~\cite{Jakubczyk_2024}, by a sequence of exact transformations, the partition function of the model may be cast in the following form  
\begin{equation} 
\label{Xi1}
\Xi(T,V,\mu_1,\mu_2)=-\frac{\beta V}{2\pi\sqrt{a_1' a_2'}}\int_{\alpha_1-i\infty}^{{\alpha_1+i\infty}}dt_1 \int_{\alpha_2-i\infty}^{{\alpha_2+i\infty}}dt_2 e^{-V\Phi(t_1,t_2)}\;, 
\end{equation}
valid for $D:=a_1 a_2-a_{12}^2>0$, and  
\begin{equation} 
\label{Xi2}
    \Xi(T, V, \mu_{1}, \mu_{2}) = -\frac{ i \beta V}{2 \pi \sqrt{a'_{1} |a'_{2}}|} \int_{\alpha_{1} - i \infty}^{\alpha_{1} + i \infty} dt_{1} \int_{ -\infty}^{\infty}dt_{2} e^{-V\Phi(t_{1}, t_{2})}
\end{equation}
for $D<0$.
The function $\Phi(t_1, t_2)$ is given as 
\begin{align} 
\label{Phi_fun}
\Phi(t_1, t_2)=& -\sum_{i=1}^2\frac{\beta}{2a_i'}(t_i-\mu_i')^2-\frac{1}{\lambda_1^3}g_{5/2}(e^{\beta t_1})-\frac{1}{\lambda_2^3}g_{5/2}(e^{\beta (\frac{a_{12}}{a_1}t_1+t_2)})  \\ 
&+\frac{1}{V}\ln (1-e^{\beta t_1}) +\frac{1}{V}\ln (1-e^{\beta (\frac{a_{12}}{a_1}t_1+t_2)})\;, \nonumber
\end{align} 
where $\lambda_i=h/\sqrt{2\pi m_i k_B T}$ are the usual thermal de Broglie lengths, and we introduced $a_1'=a_1$, $a_2'=a_2(1-\frac{a_{12}^2}{a_1 a_2})$, $\mu_1'=\mu_1$, $\mu_2'=\mu_2-\frac{a_{12}}{a_1}\mu_1$. Finally, the Bose function \cite{Ziff_1977} is defined as 
$ 
 g_\alpha (x)=  \sum_{k=1}^\infty x^k/k^\alpha.
 $ 
Due to the volume factor in the argument of the exponential in Eqs.~(\ref{Xi1}, \ref{Xi2}), the saddle-point evaluation of the integral becomes exact in the thermodynamic limit $V\to\infty$. This property arises as a consequence of the mean-field form of the interaction introduced in Eq.~(\ref{imp_B_mix}). It follows that 
\begin{align}
    \omega=-\beta^{-1}V^{-1}\ln\Xi\longrightarrow \beta^{-1}\Phi (\bar{t_1},\bar{t_2})
\end{align}
in the thermodynamic limit $V\to\infty$.  The evaluation of the free energy, therefore, boils down to 
finding $\Phi(\bar{t}_1, \bar{t}_2)$, where $(\bar{t_1},\bar{t_2})$ is the dominant saddle point of $\Phi(t_1,t_2)$.

The average densities $(n_1,n_2)$ of the mixture constituents may be related to $(\bar{t}_1,\bar{t}_2)$ by the simple formulas 
\begin{align} 
    n_1 =& -\frac{1}{a_1}(\bar{t_1}-\mu_1)+\frac{1}{a_2'}(\bar{t_2}-\mu_2')\frac{a_{12}}{a_1} \,,\nonumber \\
    n_2 =& -\frac{1}{a_2'}(\bar{t_2}-\mu_2')  \;, 
    \label{tton}
\end{align} 
which follows from 
\begin{equation}
    n_i = -\frac{\partial\omega}{\partial \mu_i}=-\beta^{-1}\frac{\partial\Phi(\bar{t}_1,\bar{t}_2)}{\partial\mu_i}\;.
\end{equation}
Eqs.~(\ref{tton}) allow for recasting the saddle-point equations $\left.\partial\Phi/\partial t_1 \right|_{\{t_i=\bar{t}_i\}} = \left.\partial\Phi/\partial t_2 \right|_{\{t_i=\bar{t}_i\}}=0$ as coupled algebraic equations for the densities:
\begin{align}
n_1=&\frac{1}{\lambda_1^3}g_{3/2}(e^{\beta(\mu_1-a_1 n_1- a_{12}n_2)})+\frac{1}{V} \frac{e^{\beta(\mu_1-a_1 n_1- a_{12} n_2)}}{1-e^{\beta(\mu_1-a_1 n_1- a_{12}n_2)}}    \label{n1eq}\\ 
n_2=&\frac{1}{\lambda_2^3}g_{3/2}(e^{\beta(\mu_2-a_2 n_2- a_{12}n_1)})+\frac{1}{V} \frac{e^{\beta(\mu_2-a_2 n_2- a_{12} n_1)}}{1-e^{\beta(\mu_2-a_2 n_2- a_{12}n_1)}} \label{n2eq}\;,
\end{align}
where the terms proportional to $1/V$ represent the condensates' densities for $V\to \infty$. The criterion for determining the presence or absence of condensates is based on the behaviour of these quantities in the thermodynamic limit as $V\to\infty$. Non-zero densities of particles in the one-particle ground states indicate the existence of condensates.

Analogously, 
the function $\Phi$ given by Eq.~(\ref{Phi_fun}) can be recast as 
\begin{align} 
&\Phi(n_1,n_2)= -\frac{\beta}{2}\left[a_1 n_1^2 +a_2n_2^2 +2a_{12} n_1 n_2\right] \nonumber \\
              & -  \frac{1}{\lambda_1^3}g_{5/2}\left(e^{\beta(\mu_1-a_1 n_1- a_{12}n_2)}\right) -\frac{1}{\lambda_2^3}g_{5/2}\left(e^{\beta(\mu_2-a_2 n_2- a_{12}n_1)}\right) \nonumber \\
              & +\frac{1}{V}\ln\left(1-e^{\beta(\mu_1-a_1 n_1- a_{12}n_2)} \right) \label{Phieq} \nonumber\\
              & +\frac{1}{V}\ln\left(1-e^{\beta(\mu_2-a_2 n_2- a_{12}n_1)} \right)\;.     
\end{align}
Notice that the last two terms in the above expression always vanish in the thermodynamic limit, in contrast to their counterparts in Eqs (\ref{n1eq},\ref{n2eq}), which yield the condensate densities. The thermodynamic state is determined by solving the self-consistency equations [see Eq.~(\ref{n1eq}) and Eq.~(\ref{n2eq})] and checking whether the one-particle ground states are macroscopically occupied. When more than one self-consistent solutions exist, the physically realized state corresponds to the dominant saddle point, which is the solution where \(\Phi\) has the lowest value.
 We can use Eq.~(\ref{n1eq},\ref{n2eq}) to determine $n_{1}$ as a function of $n_{2}$ (or vice versa). 
 Within the normal phase, we have: 
    \begin{align}
        n_{1}(n_{2}) =& \frac{ \mu_{2} - a_{2} n_{2} }{a_{12}} - \frac{1}{a_{12} \beta}\ln\left[ g_{3/2}^{-1}\left(\lambda_{2}^{3} n_2\right) \right] \label{eq:n1_Normal}\;,
    \end{align} 
    while for the BEC$_2$ phase involving condensation of type-2 (but not type-1) particles, we obtain: 
    \begin{align}
     n_{1}(n_{2}) =& \frac{ \mu_{2} - a_{2} n_{2} }{a_{12}}\label{eq:n1_BEC2}\;,
\end{align}
    
As demonstrated in Ref.~\cite{Jakubczyk_2024} (see also \cite{Schaeybroeck_2013}), if the transition between the normal and BEC$_2$ phases is continuous (and, therefore,  there is only one solution of the saddle-point equations), the phase boundary in the space spanned by $(T,\mu_1, \mu_2)$ is described by the following relation: 
    \begin{equation}
        \mu_{1} = \frac{a_{1}}{a_{12}} \mu_{2} - \frac{a_{1}a_{2} - a_{12}^{2}}{a_{12}}n_{2,c} + \frac{1}{\beta} \ln \left[ g_{3/2}^{-1}\left( \frac{ \lambda_{1}^{3} }{ a_{12} } \left(\mu_{2} - a_{2} n_{2,c}\right) \right) \right]\;,\label{eq:BEC2boundary}
    \end{equation}
where  $n_{2,c} = \frac{1}{ \lambda_{2}^{3} }\zeta\left( 3/2 \right)$.

Importantly, for certain parameter ranges (see \cite{Jakubczyk_2024}), the system admits multiple solutions and identifying the physically realized equilibrium requires checking which of them corresponds to the lowest value of $\Phi$; the remaining solutions then represent metastable or unstable equilibria. Competition between the distinct solutions leads in these cases to the occurrence of a first-order phase transition, characterised by discontinuities in particle concentrations.

The previous analysis of Ref.~\cite{Jakubczyk_2024}, largely by numerical means, explored the general structure of the phase diagram in the interaction- and mass-balanced case with repulsive interactions. The major findings of that study are summarized as follows. For $D>0$, the phase diagram hosts, in addition to the normal phase, a phase involving condensate of only type-1 particles (BEC$_1$), only type-2 particles (BEC$_2$) and condensates of both species (BEC$_{12}$), where all the involved phase transitions are continuous. This situation is schematically depicted in Fig.~{\ref{Schematic_p_d} (the right panel). In contrast, for $D<0$, there is no BEC$_{12}$ phase, and the transition between BEC$_1$ and BEC$_2$ is of first order and is accompanied by phase separation; condensation of single species of particles may occur either via a first order or continuous transition depending on system parameters. In addition, for sufficiently large, positive interspecies couplings $a_{12}$, there occurs a line of liquid-gas-type phase transitions within the normal phase; for a schematic illustration, see Fig.~\ref{Schematic_p_d} (left panel). This picture was primarily obtained through numerical analysis of  Eq.~(\ref{Phi_fun},\ref{n1eq},\ref{n2eq}) for situations with neither mass nor interaction imbalance and repulsive interparticle forces. The role of the sign of \(D\) had been recognized in much earlier studies on this  topic. Specifically, at \(T=0\), varying \(D\) across zero leads to a mixing-demixing transition in the system. For temperatures above zero  with \(D=0\), the transition between the BEC\(_1\) and BEC\(_2\) phases is expected to be of a multicritical nature \cite{Jakubczyk_2024}. 

We presently systematize, generalize, and extend that analysis to parameter ranges not addressed before, also providing a much more thorough analytical understanding. We begin by discussing the possibility of attractive couplings.

\section{Attractive interspecies interactions}
We first observe that the one-component counterpart of the system given by Eq.~(\ref{imp_B_mix}) is well-defined only for non-negative values of the interaction coupling. In regard to the presently studied two-component mixture, we note that the structure of the saddle-point equations [Eq.~(\ref{n1eq},\ref{n2eq})] does not impose immediately obvious constraints on the values of the interaction couplings $\{a_1,a_2,a_{12}\}$. In fact, as we will demonstrate below, while \(a_1\) and \(a_2\) must remain positive to prevent collapse, the system can still be stable for sufficiently small negative values of \(a_{12}\), as the repulsive intraspecies interactions can overcome the impact of the attractive interspecies coupling. To identify restrictions on the values of these parameters stemming from the system stability, we reexamine the structure of the integrals of Eq.~(\ref{Xi1},\ref{Xi2}). We first point out \cite{Napiorkowski_2011, Napiorkowski_2013, Jakubczyk_2024} that the result is independent of the arbitrary parameters $\alpha_1$ and $\alpha_2$ occurring in the integration limits. These parameters are taken to be negative to guarantee that the function $\Phi (t_1,t_2)$ can (at least for repulsive interactions) be analytically continued into the region of the complex plane encompassing the integration contour. 
To ensure that the Bose function $g_{5/2}(x)$ and the logarithmic function $\ln(x)$ are well-defined, both $|e^{\beta t_{1}}|$ and $|e^{\beta (\frac{a_{12}}{a_{1}}t_{1} + t_{2})}|$ must be less than one. This condition is satisfied provided that the real parts of $t_{1}$ and $(\frac{a_{12}}{a_{1}}t_{1} + t_{2})$ are negative. When $D >0$, it is straightforward to see that this can be guaranteed by appropriate choice of $\alpha_{1}$ and $\alpha_{2}$ (irrespective of the sign of $a_{12}$). On the other hand, when $D < 0$ and $a_{12}<0$, it is impossible to fulfil the above conditions. To see this, we express $t_{1}$ as the sum of its real and imaginary parts, $t_{1} = a + ib$. Fulfillment of the inequality $|e^{\beta t_{1}}|< 1$ requires that $a<0$. However, the condition $|e^{\beta (\frac{a_{12}}{a_{1}}t_{1} + t_{2})}|< 1$ requires $a$ to be positive, as $a_{12}<0$ and $t_{2}$ is integrated over the entire real axis. This leads to a contradiction, rendering it impossible to satisfy both conditions simultaneously when $D<0$. In consequence, the integrals in Eq.~(\ref{Xi1}, \ref{Xi2}, \ref{Phi_fun}) remain well defined also for attractive interspecies interactions but only as long as $D>0$. We also note that the intraspecies couplings $a_1$, $a_2$ must necessarily be positive.

It is now interesting to examine the structure of the phase diagram, as the interspecies coupling $a_{12}$ is decreased starting from a small positive value, crosses zero, and approaches the instability limit, where $D\to 0^+$. This is illustrated in Fig.~\ref{fig:attractive}, where we plot a projection of the phase diagram on the $(\mu_1,\mu_2)$ plane for a sequence of values of $a_{12}$ (with $a_1=a_2$ and temperature $T$ fixed). For the present case (where $D>0$), there exists only one solution of the saddle-point equations and the normal-BEC$_2$ phase boundary may be extracted using Eq.~(\ref{eq:BEC2boundary}) [and the normal-BEC$_1$ phase boundary analogously]. This fact is not obvious at all and stands in contrast to the case $D<0$. We will discuss and demonstrate this in detail in Sec.~IV.
\begin{figure*}
    \centering
    \includegraphics[width=0.49\textwidth]{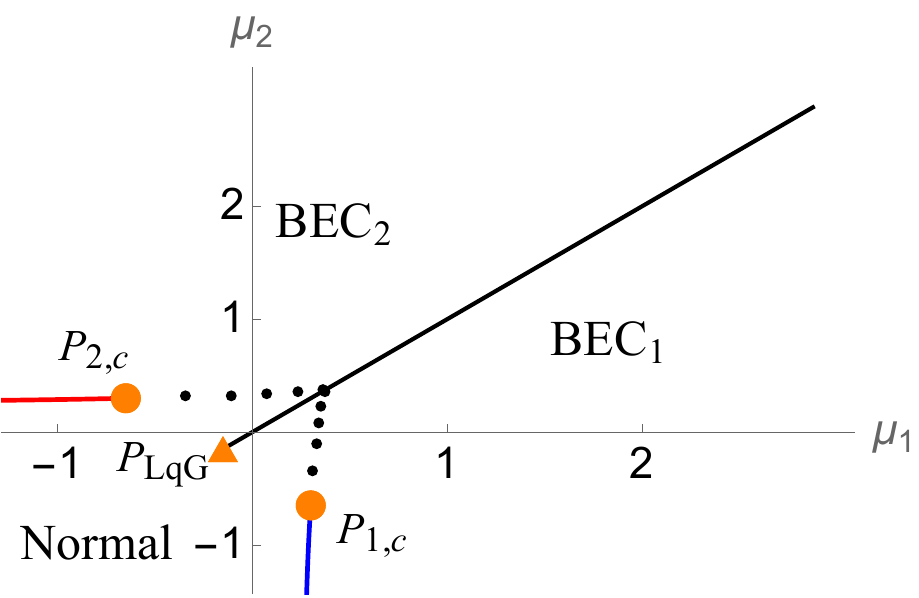}
   \includegraphics[width=0.49\textwidth]{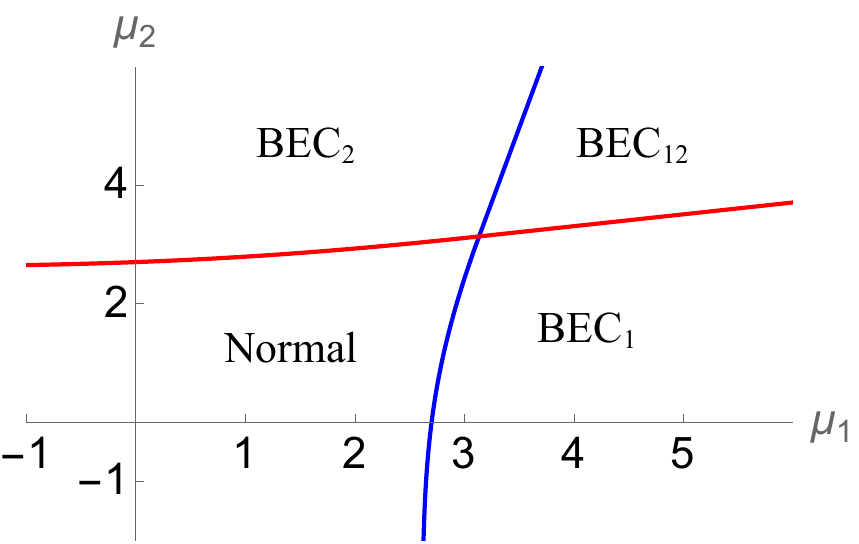}
    \caption{Schematic illustration of projections of the phase diagram on the $(\mu_1,\mu_2)$ planes (fixed $T$) - see the main text. For $D>0$ (right panel), all the involved phase transitions are continuous and the system always hosts a quadruple point, where the four thermodynamic phases coexist. In contrast, for $D<0$ (left panel), the BEC$_{12}$ phase is absent, the condensation may, but does not have to, occur via a first-order transition (thus implying the appearance of tricritical points (denoted here as $P_{1,c}$ and $P_{2,c}$). In addition, a liquid-gas type phase transition may (but does not have to) appear, which terminates with a critical point ($P_{LqG}$). The BEC$_1$-BEC$_2$ phase transition is first order, and the transition between the normal and BEC$_1$ (or BEC$_2$) phases is first order (dotted black lines) or second order (full lines).}  
    \label{Schematic_p_d}
\end{figure*}

\begin{figure}
    \centering
    \includegraphics[width=1\linewidth]{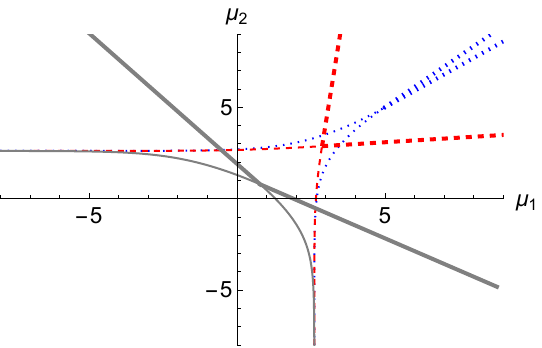}
    \caption{Projection of the phase diagram on the $(\mu_{1},\mu_{2})$ plane for a sequence of values of the interspecies coupling $a_{12}$ [$\beta a_{12}\lambda_1^{-3} =0.9$  (blue dotted lines), $\beta a_{12}\lambda_1^{-3}=0.1$ (red dashed lines), and $\beta  a_{12}\lambda^{-3}_1=-0.7$ (grey lines)] for sufficiently weak interspecies coupling $a_{12}$ such that $D>0$. In each of the plotted cases the general structure of the phase diagram is analogous to the one of the right diagram of Fig.~1;  the BEC$_{12}$ phase occupies the wedge-shaped region characterized by large values of $\mu_1$ and $\mu_2$ (corresponding also to high densities). The system hosts a quadruple point, where the BEC$_{12}$, BEC$_1$, BEC$_2$ and the normal phase coexist. For $a_{12}>0$ and $D=0$ the BEC$_{12}$ phase does not exist (the corresponding wedge angle $\phi=0$); the BEC$_{12}$ wedge opens upon decreasing $a_{12}$, its angle reaches the value $\phi=90$\degree\, for $a_{12}=0$ and approaches $\phi=180$\degree\, in the  limit $a_{12}\to -\sqrt{a_1 a_2}^+$ (where $D\to0^+$ and the system loses its stability). The plot parameters are $\beta a_1\lambda_1^{-3}= \beta a_2\lambda_1^{-3}=\lambda_1/\lambda_2=1$. All the phase transitions are continuous (see Sec.~IV).}
    \label{fig:attractive}
\end{figure} 
For $D>0$ and relatively small and positive values of $a_{12}$, the BEC$_{12}$ phase is stable in the acute wedge-shaped region characterized by large values of $\mu_1$ and $\mu_2$ (which also correspond to high densities) - see Fig.~\ref{fig:attractive}. The system hosts a quadruple point, where the BEC$_{12}$ phase coexists with the BEC$_1$, BEC$_2$, as well as with the normal phase. All the phase transitions involved here are continuous (see Sec.~IV). Upon reducing $a_{12}$, the position of the quadruple point becomes shifted to lower values of $(\mu_1, \mu_2)$, and the BEC$_{12}$ wedge progressively opens. The wedge angle $\phi$ reaches the value 90\degree$\,$ for $a_{12}=0$ and is further increased as $a_{12}$ becomes negative. The limiting value $a_{12}\to -\sqrt{a_1a_2}^+$ (where $D\to0^+$) corresponds to the fully open wedge  ($\phi\to$180\degree). This picture may  be understood realizing  that at the BEC$_2$-BEC$_{12}$ transition one has [see Eq.~(\ref{n1eq}, \ref{n2eq})]:
\begin{equation}
\mu_1-a_1 n_1-a_{12}n_2=0\,,\;\;\;\mu_2-a_2 n_2-a_{12}n_1=0\,,\;\;\; n_1=n_{1,c} 
\end{equation}
and analogously for the BEC$_1$-BEC$_{12}$ line. This allows us to express the BEC$_2$-BEC$_{12}$ and BEC$_1$-BEC$_{12}$ phase boundaries as 
\begin{equation}
\mu_2 -\frac{a_2}{a_{12}}\mu_1+D\frac{n_{1,c}}{a_{12}}=0    
\label{Eqqq1}
\end{equation} 
and 
\begin{equation}
\mu_1 -\frac{a_1}{a_{12}}\mu_2+D\frac{n_{2,c}}{a_{12}}=0\;, 
\label{Eqqq2}    
\end{equation} 
respectively. For $D\to 0^+$ these two lines coincide. 

The location of the line of quadruple points in the $(T,\mu_1,\mu_2)$ space follows immediately from Eq.~(\ref{Eqqq1},\ref{Eqqq2}) and is given by 
\begin{equation}
\mu_2^{(Q)}(T)=\zeta\left(3/2\right)\frac{(2\pi  k_B T)^{3/2}}{h^3}\left(a_2 m_2^{3/2}+a_{12}m_1^{3/2}\right)    
\end{equation}
and analogously for $\mu_1^{(Q)}(T)$. The line exists for all temperatures. 
Note that the sign of $\mu_2^{(Q)}(T)$ may be both positive and negative but cannot be changed by varying $T$. For the specific choice  $a_2= - a_{12}(m_1/m_2)^{3/2}$ $\mu_2^{(Q)}(T)$ vanishes for all $T$.

It is remarkable that the saddle-point equations are 'unaware' of the instability occurring for $a_{12}< -\sqrt{a_1a_2}$, i.e. one might naively plot the phase diagram also for this case. The relevant information concerning the stability region in the parameter space is, on the other hand, easily accessible at the level of the integrals in Eq.~(\ref{Xi1}, \ref{Xi2}), which are not defined for $D<0$ and $a_{12}<0$. 

The analysis of attractive interspecies interactions concludes here; henceforth, we will consider \(a_{12} > 0\), where additional structures of the phase diagram appear.

\section{Tricritical points for condensation} 
In this section, we analyse the conditions for realising condensation as a first-order transition. Characterised by discontinuities of $n_1$, and $n_n$ as functions of $\mu_1$, $\mu_2$ and $T$. We demonstrate the impossibility of such a scenario for $D>0$ (a fact used in the analysis of the previous section). We also show that $D<0$ is a necessary but not sufficient condition for the appearance of first-order condensation in the system's phase diagram. This scenario will, however, always be realised for $T$ sufficiently low.  

We focus on the transition line between the normal and BEC$_2$ phases; the normal-BEC$_1$ case may be addressed in a fully analogous manner. For sufficiently large negative values of $\mu_1$ ($-\beta\mu_1 \gg 1$), the concentration of type-1 particles is negligibly low and properties of the mixture are equivalent to those of the one-component system. In particular, there is only one solution to the saddle-point equations;  condensation of type-2 particles occurs via a continuous transition, and the corresponding phase boundary follows Eq.~(\ref{eq:BEC2boundary}). It may, however, happen \cite{Jakubczyk_2024} that, as one moves along the line defined by Eq.~(\ref{eq:BEC2boundary}), additional solutions to the saddle-point equation appear and Eq.~(\ref{eq:BEC2boundary}) begins to represent an unstable equilibrium. This scenario requires an appearance of a tricritical point in the phase diagram ($P_{2,c}$ in Fig.~1), which is reflected by a change of the character of the extremum of $\Phi(n_{1}(n_{2}), n_{2} )$ at the critical density. This, in turn, implies an existence of a point on the line given by Eq.(\ref{eq:BEC2boundary}) where the second derivative of $\Phi(n_{1}(n_{2}), n_{2} )$ with respect to $n_{2}$ vanishes. Using Eq.~(\ref{eq:n1_BEC2}), we view $\Phi(n_{1} , n_{2} )$, as a function of $n_{2}$ only and we introduce  $\Tilde{\Phi}(n_{2})\equiv\Phi(n_{1}(n_{2}) , n_{2} )$. The condition for vanishing of the second derivative of $\Tilde{\Phi}(n_{2})$ then may be written as:
    \begin{equation}
        g_{1/2}\left[ e^{\beta( \mu_{1,tri} - \frac{a_{1}}{a_{12}} \mu_{2,tri} +\frac{ D }{ a_{12} } n_{2,c})} \right] = \frac{a_{2}\lambda_{1}^{3}}{-D\beta }\;. \label{eq:mu2c}
    \end{equation} 
Since the left-hand side of the above relation is nonnegative, it immediately follows that a tricritical point in question may never exist for $D>0$. Therefore, the transition between the normal and BEC$_2$ phases is bound to be continuous in this case. We therefore now restrict our analysis to $D<0$ and recall that the coordinates of the tricritical point $\mu_{1,tri}$ and $\mu_{2,tri}$ are constrained by Eq.~(\ref{eq:BEC2boundary}), which allows for evaluating them as:
\begin{align} 
\label{tric1}
{\mu}_{1,tri} =& a_{12} n_{2,c} + \frac{a_{1}}{\lambda_{1}^{3}} g_{3/2}\left[ g_{1/2}^{-1}\left( \frac{ a_{2}\lambda_{1}^{3} }{-D\beta} \right) \right] + \frac{1}{\beta} \ln \left[ g_{1/2}^{-1}\left( \frac{ a_{2}\lambda_{1}^{3} }{-D\beta} \right) \right]\;,\\ 
\label{tric2}
        {\mu}_{2,tri} =& a_{2} n_{2,c} + \frac{a_{12}}{\lambda_{1}^{3}} g_{3/2}\left[ g_{1/2}^{-1}\left( \frac{ a_{2}\lambda_{1}^{3} }{-D\beta } \right) \right]\;.
    \end{align} 
As already remarked, the tricritical point corresponding to the phase boundary between the normal and BEC$_1$ phases can be analyzed in a fully analogous manner.      

We emphasize that the above discussion implies that the tricritical points may occur exclusively for negative values of the determinant $D$. However, a negative $D$ does not necessarily imply that Eqs.~(\ref{tric1}, \ref{tric2}) indeed represent a meaningful tricritical point on the phase diagram. The additional requirement is that the putative tricritical point of Eq.~(\ref{tric1}, \ref{tric2}) be located to the left (i.e. correspond to lower values of the chemical potentials) from the point of intersection of the putative second-order phase boundaries [given by Eq.~(\ref{eq:BEC2boundary}) and an analogous equation for the normal-BEC$_1$ transition]. Numerical analysis indicates that this condition becomes fulfilled for sufficiently low temperatures and/or $D$ sufficiently close to zero. This is easily proven in the balanced case, where $a_1=a_2=a$ and $\lambda_1=\lambda_2=\lambda$. In this case, the condition for Eq.~(\ref{tric1}, \ref{tric2}) to represent a physical tricritical point reduces to $\mu_{1,tri}<\mu_{2,tri}$. After applying Eqs.~(\ref{tric1}, \ref{tric2}) and recalling that $n_{2,c} = \frac{1}{ \lambda_{2}^{3} }\zeta\left( 3/2 \right)$, this condition  may be recast as:
\begin{equation}
\left(a_{12}-a\right)\left[\zeta(3/2)-g_{3/2}(\mathcal{X})\right] +\lambda^3\beta^{-1}\ln(\mathcal{X})<0\;,   
\label{ineq}
\end{equation}
where we introduced $\mathcal{X}:= g_{1/2}^{-1}\left(a\lambda^3/(-D\beta)\right)$. We observe that $\mathcal{X}\in (0,1)$, which implies that the second term on the left-hand side of Eq.~(\ref{ineq}) is always negative, in contrast to the manifestly positive first term. 
We note that $a\lambda^3/(-D\beta)\sim 1/(|D|\sqrt{T})$ and the second term always dominates in the low-$T$ limit, such that Eq.~(\ref{ineq}) remains fulfilled. 
On the other hand, the condition of Eq.~(\ref{ineq}) is always violated in the limit of high temperatures [since the first term approaches $(a_{12}-a)\zeta(3/2)>0$ and the second term scales to zero as $\sim \ln(T)/\sqrt{T}$. The same happens in the asymptotic regime of large $(a_{12}-a)$, where the first term in Eq.~(\ref{ineq}) is positive and scales linearly with $(a_{12}-a)$, while the second term remains negative, but diverges only logarithmically for $(a_{12}-a)$ large. 
The evolution of positions of the (putative) tricritical points, indicated by orange discs, upon varying the system parameters is illustrated in Figs.\ref{fig:cr_int} and \ref{fig:cr_temp}. 

The conclusions of this section are summarized as follows: (i) tricritical points on the normal-BEC phase boundary (and in consequence also first-order condensation) may never occur for $D>0$; (ii) for $D<0$ they are always present on a projection on the phase diagram on the plane spanned by $(\mu_1, \mu_2)$ provided $T$ or $|D|$ is sufficiently low; (iii) they never occur for $T$ or $|D|$ sufficiently high.

\section{Liquid-gas type transition within the normal phase}
As reported in the recent study of Ref.~\cite{Jakubczyk_2024}, for sufficiently large, positive values of $a_{12}$, the two-component Bose mixture may undergo a phase transition which does not involve condensation, but manifests itself in a jump of concentrations of the mixture constituents (both remaining below the critical condensation density). There is no fundamental difference between the two non-condensed phases. This transition, characterized by a jump in densities, is analogous to the liquid-gas transitions occurring in classical fluids. Interestingly, a somewhat similar transition was identified \cite{Spada_2023} via Monte Carlo simulations of a two-dimensional Bose mixture in a completely different parameter regime characterized by attractive interspecies couplings. Below, we demonstrate that within the present mean-field model in dimensionality $d=3$, the occurrence of this transition is completely excluded as long as $D>0$. We identify the parameter ranges (corresponding to sufficiently large positive $a_{12}$ or $T$), where it does occur in the case $D<0$. We also find that there exists a range of system parameters where the two normal phases coexist with the two BEC phases, implying the presence of quadruple points.

Anticipating the possibility of first-order transitions terminating at critical points, and occurring within the non-condensed region of the phase diagram, we employ a method analogous to the approach of the previous section with the aim of locating the critical points. We express $n_{1}$ as a function of $n_{2}$ using Eq.~(\ref{eq:n1_Normal}), which enables us to define $\Tilde{\Phi}(n_{2})$. At the critical point, the second derivative of $\Tilde{\Phi}(n_{2})$ must vanish. Unlike the case analyzed in the previous section, we do not have an explicit formula for the coexistence line. In the general situation, the missing equation is provided by the condition of vanishing third derivative of $\Tilde{\Phi}(n_{2})$. The set of equations to tackle the problem involves two equations corresponding to the saddle point conditions and two conditions involving higher derivatives: $\Tilde{\Phi}''(n_{2})=\Tilde{\Phi}'''(n_{2})=0$. By elementary algebraic manipulations, the condition $\tilde{\Phi}''(n_{2}) = 0$ may be cast in the following form:
 \begin{equation} 
 \label{Phibis}
    \left(D \beta  y +a_{1}\lambda_{2}^3\right) \left(D\beta ^2 x y+ a_{1}\lambda_{2}^3 \beta  x + a_{2} \lambda_{1}^3\beta  y +\lambda_{1}^3\lambda_{2}^3\right) = 0, 
\end{equation}
where we introduced
\begin{equation}
    \begin{aligned}
        x =& g_{1/2}\left[ e^{\beta \left( \mu_{1} - a_{1} n_{1} - a_{12} n_{2} \right)} \right]\;, \\
        y =& g_{1/2}\left[ e^{\beta \left( \mu_{2} - a_{2} n_{2} - a_{12} n_{1} \right)}\right]\;. 
    \end{aligned} 
\end{equation}
It is immediately clear that Eq.~(\ref{Phibis}) is never satisfied for \(D > 0\). This implies that any critical point in normal phases (or on its boundary) can only arise for $D<0$. However, a negative $D$ does not necessarily imply that solutions of Eq.~(\ref{Phibis}) represent meaningful critical points in the phase diagram.

In the general case, the relevant set of equations can be studied numerically. Significant simplifications occur in the symmetric case  (\( a_{1} = a_{2} = a \) and \( \lambda_{1} = \lambda_{2} = \lambda \)) in which the system of equations can be solved, leading to the following expression for the location of the critical point \( P_{LqG}=(\mu_{c},\mu_{c}) \):
    \begin{equation}
        \mu_{c} = \frac{(a + a_{12})}{\lambda^{3}}g_{3/2}\left[ g_{1/2}^{-1}\left( \frac{\lambda^{3}}{\beta(a_{12}-a)} \right) \right] + \frac{1}{\beta}\ln \left[ g_{1/2}^{-1}\left( \frac{\lambda^{3}}{\beta(a_{12}-a)} \right)\right]. \label{eq:muc}
    \end{equation}
We also obtain an expression for the density $n_{LqG}$ of each of the mixture constituents at the critical point :
    \begin{equation}
        n_{LqG}=\frac{1}{\lambda^{3}}g_{3/2}\left[ g_{1/2}^{-1}\left( \frac{\lambda^{3}}{\beta(a_{12}-a)} \right) \right].\label{eq:nlqg}
    \end{equation}
A liquid-gas-type phase transition occurs if $P_{LqG}$ is lower than the triple point $P_{T}$. In this case, in a projection of the phase diagram on a fixed $T$ plane, a straight line of first-order phase transitions is present and connects the triple and the critical points.

 Let us now inspect the regime of large \( a_{12} \) or \( T \). In this case, the triple points involving the BEC do not exist (see Sec.~IV), and the BEC$_1$, BEC$_2$ and normal phases meet at the intersection of the second-order transition lines (see Sec.~II) at \( P_{T}=(\mu_{T},\mu_{T}) \). In the present limit, \( \mu_{c} \) approaches minus infinity. To analyze \( \mu_{T} \), we recast Eq.~(\ref{eq:BEC2boundary}) as:
    \begin{equation}
        \left(a_{12}-a \right)\mu_{T} =  \left( a_{12}^{2} - a^{2}\right)\frac{\zeta\left( 3/2 \right)}{\lambda^{3}} + \frac{a_{12}}{\beta} \ln \left[ g_{3/2}^{-1}\left( \frac{ \lambda^{3} }{ a_{12} } (\mu_{T} - a \frac{\zeta\left( 3/2\right)}{\lambda^{3}} \right) \right]\;,
    \end{equation}
which shows that $\mu_{T}$ approaches positive infinity in the present regime. We conclude that the liquid-gas-type transition exists for sufficiently large $a_{12}$ or  $T$. The evolution of the positions of the (presumed) critical point, indicated by a triangle, is shown in Figs. \ref{fig:cr_int} and \ref{fig:cr_temp} as the system parameters change.

Our findings of the present section can be summarized as follows: (i) a liquid-gas-type phase transition may never occur for \(D > 0\); (ii) for \(D < 0\), it is present on the phase diagram projection on the plane spanned by \((\mu_{1}, \mu_{2})\), provided that \(a_{12}\) or \(T\) are sufficiently large; (iii) the phase transition does not occur when \(a_{12}\) or \(T\) are sufficiently low; (iv) it is possible to adjust \(a_{12}\) or \(T\) in such a way that all critical points are present on phase diagram, indicating a quadruple point; (v) for sufficiently large \(a_{12}\) or \(T\) the triple point, becomes degenerate, consisting only two phases: \(BEC_{1}\) and \(BEC_{2}\).

\begin{figure*}
    \centering
    \includegraphics[width=0.24\textwidth]{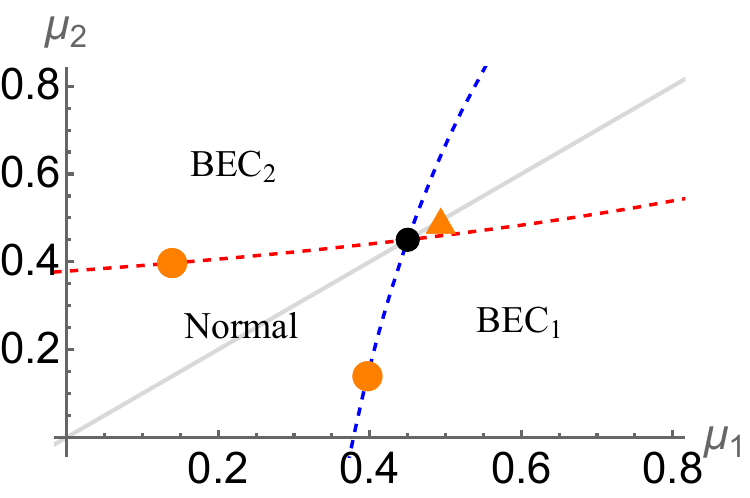}
    \includegraphics[width=0.25\textwidth]{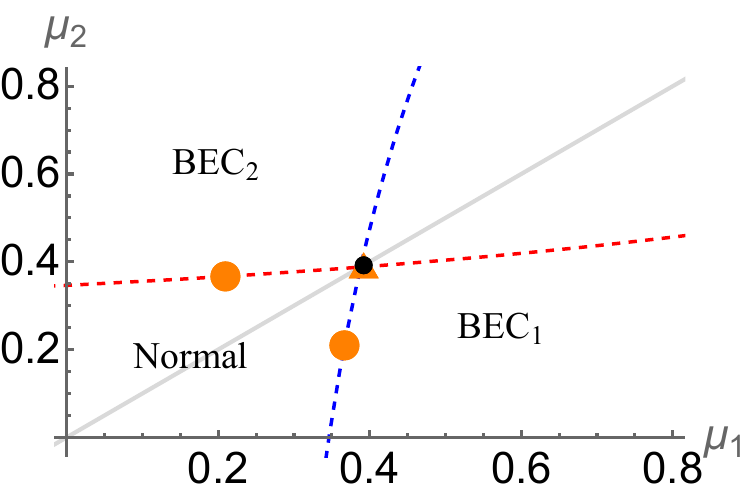} 
    \includegraphics[width=0.25\textwidth]{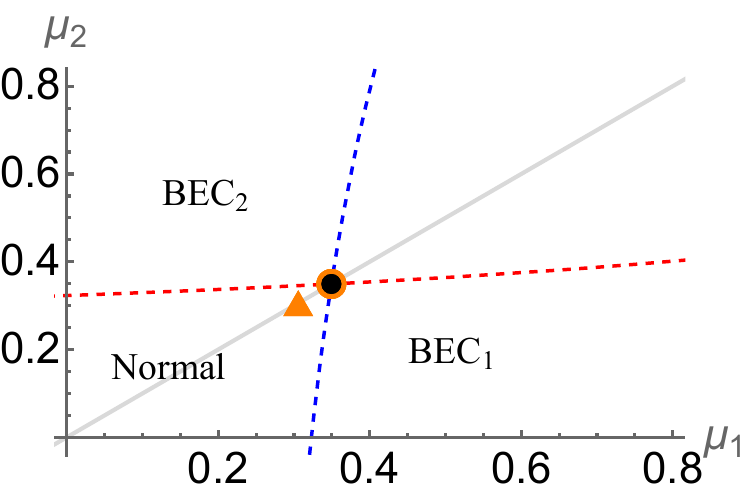}
    \includegraphics[width=0.24\textwidth]{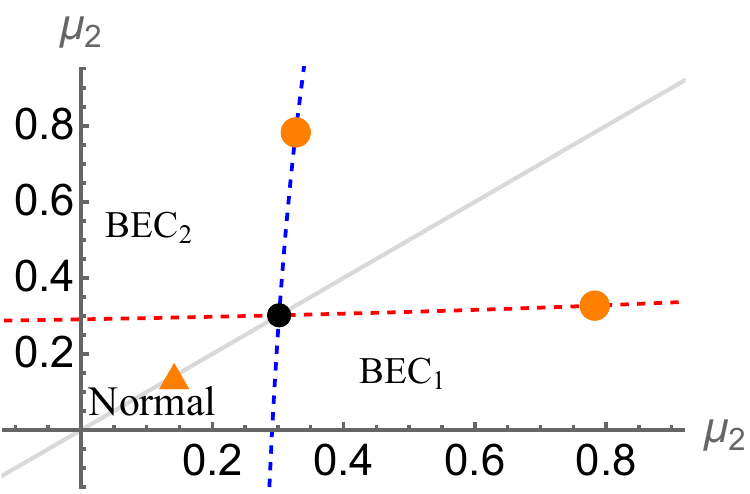}
    \caption{Evolution of the tricritical points and the critical point upon varying $a_{12}$ at a fixed temperature ($\lambda_1=\lambda_2=\lambda$, $a_1=a_2=a$, $a \lambda^{-3} =0.1$). The subsequent illustrations correspond to $\beta a_{12} \lambda^{-3} = 0.7$, $\beta a_{12} \lambda^{-3} = 0.92$, $\beta a_{12} \lambda^{-3} = 1.1$, and $\beta a_{12} \lambda^{-3}=1.5$. 
      The orange discs represent the (putative) tricritical points, the orange triangles mark the (putative) critical points, and the black discs indicate the triple points. The orange points are physical if they are located to the left (and below) of the triple point. }
    \label{fig:cr_int}
\end{figure*}

\begin{figure*}
    \centering
    \includegraphics[width=0.24\textwidth]{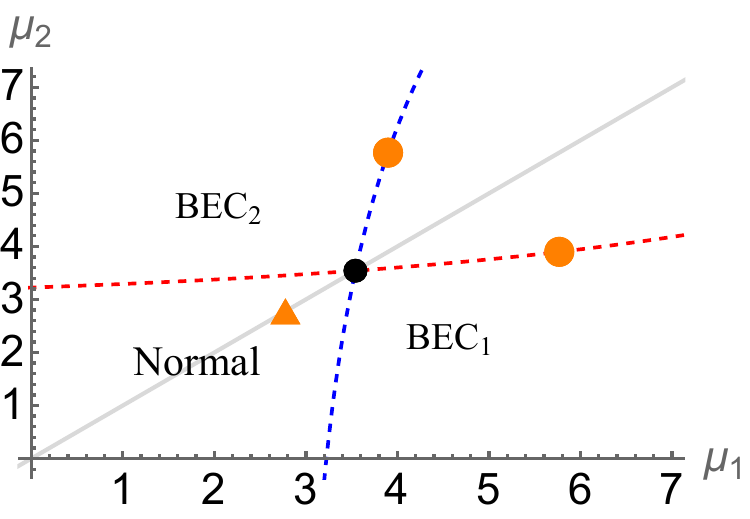}
    \includegraphics[width=0.24\textwidth]{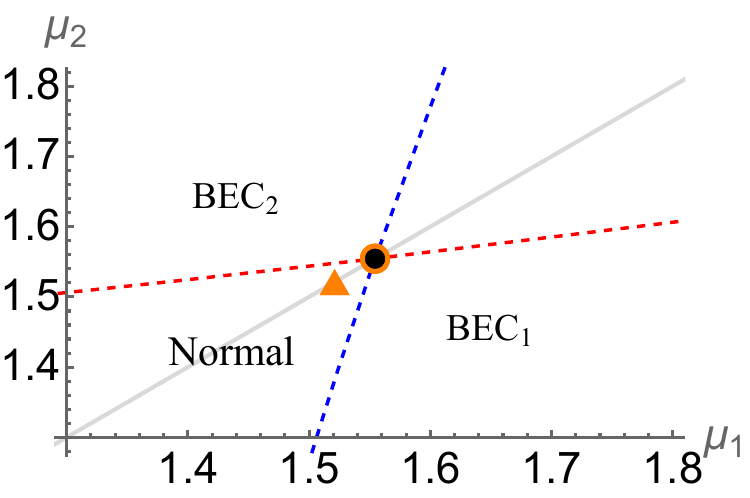}
    \includegraphics[width=0.24\textwidth]{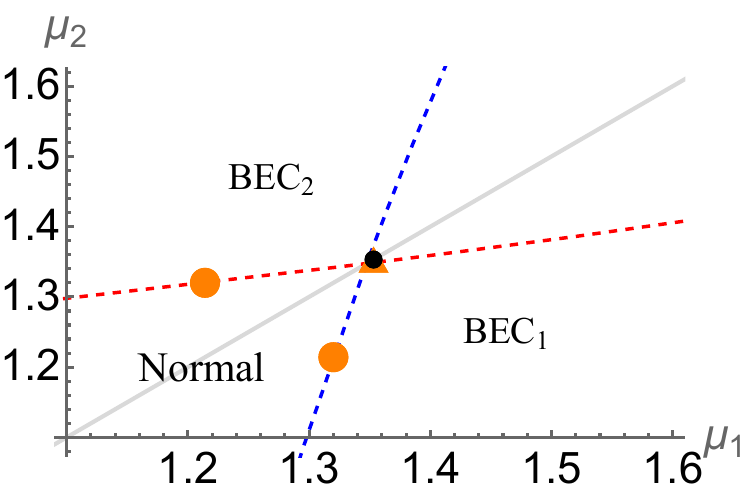}
    \includegraphics[width=0.24\textwidth]{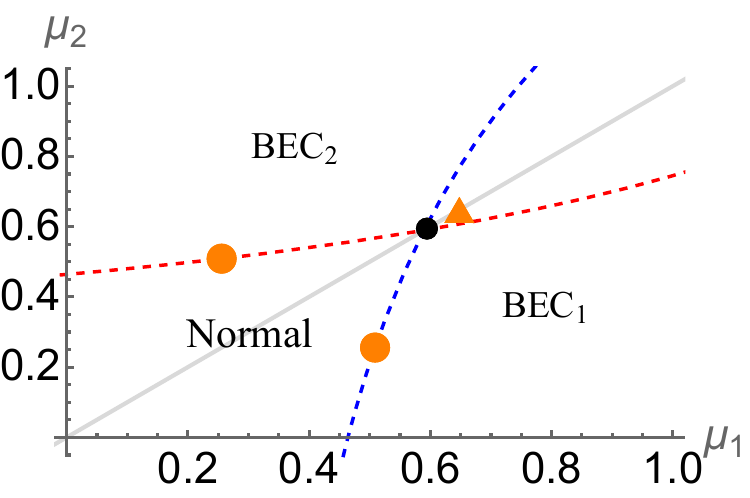}
    \caption{Evolution of the tricritical points and the critical point upon varying temperature at fixed $\frac{a_{12}}{a}=5$ ($\lambda_1=\lambda_2=\lambda$, $a_1=a_2=a$). The subsequent illustrations correspond to $\beta a \lambda^{-3} = 0.22$, $\beta a \lambda^{-3} \approx 0.16$, $\; \beta a \lambda^{-3} \approx 0.15$, and $\beta a \lambda^{-3} = 0.11$. The orange discs represent the (putative) tricritical points, the orange triangles mark the (putative) critical points, and the black discs indicate the triple points. The orange points are physical if they are located to the left (and below) of the triple point.}
    \label{fig:cr_temp}
\end{figure*}

\section{Triple points and the BEC$_{12}$ phase} 
In this section, we demonstrate in full generality the fact that, at most, one triple point may be present in the projection of the phase diagram on the $(\mu_1, \mu_2)$ plane. This is at odds with the phase diagram proposed in \cite{Schaeybroeck_2013} for $D>0$, where the system exhibits (at fixed $T$) two such triple points. We additionally demonstrate the related general fact that the BEC$_{12}$ phase may never emerge for $D<0$. We address the two cases ($D<0$ and $D>0$) separately.
\subsection{$D <0$}
We begin by showing that two triple points cannot exist when \( D < 0 \). This fact is rooted in the generic absence of the BEC$_{12}$ phase, which is demonstrated below. Our reasoning is as follows:

In the BEC$_{2}$  phase, the saddle-point equations simplify to:

\begin{equation}
\frac{ \mu_{2} - a_{2} n_{2} }{ a_{12} } = \frac{1}{\lambda_1^3} g_{3/2} \left[ e^{\beta \left( \mu_{1} - \frac{a_{1}}{a_{12}} \mu_{2} + \frac{D}{a_{12}} n_{2} \right)} \right].
\end{equation}

The right-hand side of this equation is bounded. Consequently, in the limit where \( \mu_{2} \rightarrow \infty \), we have \( n_{2} \sim \)  \( \frac{\mu_{2}}{a_{2}} \). To ensure that there is no condensation of type-1 particles in this limit, the following inequality must hold: 

\begin{equation}
\mu_{1} - \frac{a_{1}}{a_{12}} \mu_{2} + \frac{D}{a_{12}} n_{2} < 0,
\end{equation}

which for $\mu_2\to\infty$  simplifies to:

\begin{equation}
\mu_{1} - \frac{a_{12}}{a_{2}} \mu_{2} < 0,
\end{equation}
Here, we have used the relation $D = a_1 a_2 -a_{12}^2$ along with the fact that in  aforementioned limit, \( n_{2} \sim \frac{\mu_{2}}{a_{2}} \). If \( \frac{a_{2}}{a_{12}} \mu_{1} > \mu_{2} \), the BEC$_{1}$  phase occurs alongside BEC$_{2}$. In the complementary case, we find that when \( \mu_{2} > \frac{a_{12}}{a_{1}} \mu_{1} \), type-2 particles condense along type-1 particles. 

Collecting these results, we conclude that for the BEC$_{12}$  phase to occur, the chemical potentials must satisfy the following inequalities:

\begin{equation}
\frac{a_{12}}{a_{1}} \mu_{1} < \mu_{2} < \frac{a_{2}}{a_{12}} \mu_{1}\;,
\end{equation}
 which leads to the necessary condition: $a_{1} a_{2} - a_{12}^{2} > 0$. This demonstrates that generically the BEC$_{12}$  phase can only exist for \( D > 0 \). 
\subsection{$D>0$}
We first recall that for \( D > 0 \), all the phase transitions in question are continuous, and, in particular, the phase boundary between the normal and BEC$_2$ phases is given by Eq.~(\ref{eq:BEC2boundary}). We now analyze the possibility of coexistence between the different phases of the system by comparing the corresponding values of \( \Phi(n_{1}, n_{2}) \). 
 The transition lines meet at the point:
        \begin{equation}
        \begin{aligned}
            \mu_{1} =& \left( \frac{ a_{1} }{\lambda_{1}^{3}} + \frac{ a_{12} }{\lambda_{2}^{3}} \right) \zeta(3/2),\\
            \mu_{2} =&\left( \frac{ a_{2} }{\lambda_{2}^{3}} + \frac{ a_{12} }{\lambda_{1}^{3}} \right) \zeta(3/2).
        \end{aligned}
        \end{equation}
        The extremum of \( \tilde{\Phi}(n_{i}) \) along the \( BEC_{i} \) boundary line occurs at \( n_{i} = n_{i,c} \).
        It is straightforward to compute the complementary densities at this point of convergence. Specifically, the density $n_{1}$ in BEC$_{2}$ phase is given by: $n_{1} = \frac{1}{\lambda_{1}^3} \zeta(3/2)$, and the density of $n_{2}$ in BEC$_{1}$ phase is $n_{2} = \frac{1}{\lambda_{2}^3} \zeta(3/2)$. We can also compute the densities corresponding to the $BEC_{12}$ phase:
        \begin{equation}
        \begin{aligned}
            n_{1} = \frac{\mu_{1} a_{2} - \mu_{2} a_{12}}{ a_{1}a_{2} - a_{12}^{2} } = \frac{1}{\lambda_{1}^3} \zeta(3/2),\\
            n_{2} = \frac{ \mu_{2} a_{1} - \mu_{1} a_{12} }{ a_{1}a_{2} - a_{12}^{2} } = \frac{1}{\lambda_{2}^3} \zeta(3/2).
        \end{aligned}
        \end{equation}
    From this we obtain:
    \begin{equation}
        \Phi_{ \text{BEC}_{1} } = \Phi_{ \text{BEC}_{2} } = \Phi_{ \text{BEC}_{12} }.
    \end{equation}
    This implies that the BEC$_{1}$ - BEC$_{2}$ transition line may not exist.

    Our findings regarding triple points and the BEC$_{12}$ phase can be summarized as follows: (i) The BEC$_{12}$ phase does not emerge for $D < 0$. (ii) Consequently, the absence of the BEC$_{12}$ phase for $D < 0$ implies that no triple points are present in the phase diagram within this interaction regime. (iii) For $D > 0$, all phase transitions are continuous. (iv) The BEC$_{1}$–BEC$_{2}$ transition line is absent for $D > 0$.

\section{Role of interaction and mass imbalance}  
In this section, we study the impact of mass and interaction imbalance on the system. We analyze two parameter regimes. In the first one, we consider small, negative \(D\), where two tricritical points exist, while the critical point associated with the liquid-gas type phase transition is absent. We explore how the imbalance influences the tricritical points. In the second analyzed case, we consider large negative \(D\). Here, the only remaining critical point is the one related to the liquid-gas phase transition. Our discussion will focus on how this transition evolves under variation of imbalance.
\subsection{Tricritical points $P_{1,c}$ and $P_{2,c}$}

We first focus on the small-$D$ regime, fix the values of \(a_{12}\) and \(\beta\), and define the imbalance parameters \(\delta = \frac{a_1}{a_2} \geq 1\) and \(\kappa = \frac{\lambda_2}{\lambda_1} \geq 1\). As the imbalance parameters increase, the first-order phase transition between the normal and \(BEC_1\) phases becomes progressively suppressed. At sufficiently large imbalances, this first-order transition is entirely replaced by a continuous transition. However, when the temperature is lowered, the first-order transition reemerges.

The question we now address is whether the two tricritical points still exist at certain temperatures even under conditions of large imbalance.

For the first-order phase transition between the normal phase and BEC$_{1}$ to be absent, it is necessary that the point \(P_{1,c} = (\tilde{\mu}_{1, tri}, \tilde{\mu}_{2, tri})\) lies to the left of the putative boundary of BEC$_{2}$ defined by relationship \(\mu_{1}(\mu_{2})\) [see Equation: (\ref{eq:BEC2boundary})]. This condition translates into the inequality: \(\mu_{1}(\tilde{\mu}_{2, tri}) - \tilde{\mu}_{1, tri} > 0\). From earlier discussions, it is evident that, in the symmetric case, tricritical points emerge at sufficiently low $T$ or $D$. Consequently, we analyze the above inequality in the limit as $\beta \rightarrow \infty$ or $D \rightarrow 0$ while keeping $a_{1}$ and $a_{2}$ fixed. We introduce the variable $\mathcal{Y}:=a_{1}\lambda_{2}^3/\left(-D\beta\right)$. In both limits, $\mathcal{Y} \rightarrow \infty$ and we utilize the following expansion for $g_{\alpha}(x)$ around $x=1$ as derived in \cite{Ziff_1977}:
\begin{equation}
g_{\alpha}(x) = \Gamma(1-\alpha)\left( - \ln{(x)} \right)^{\alpha-1} + \sum_{k=0}^{\infty} \zeta(\alpha - k) \frac{ \left( \ln{x} \right)^{k} }{k!}.
\end{equation}
For $\alpha = 3/2$ we have:
\begin{equation}
g_{3/2}(x) \sim \zeta(3/2) + \Gamma(-1/2)\sqrt{ - \ln{(x)} }, \quad x \rightarrow 1^{-},
\end{equation}
\begin{equation}
g_{3/2}^{-1}(y) \sim \exp\left[ -\frac{\left( y-\zeta(3/2) \right)^{2}}{\Gamma\left(-1/2\right)^{2}} \right],\quad  y \rightarrow \zeta(3/2)^{-}\;,
\end{equation}
while for $\alpha = 1/2$ we obtain,
\begin{equation}
g_{1/2}(x) \sim \frac{\Gamma(1/2)}{\sqrt{ - \ln{(x)} } }, \quad x \rightarrow 1^{-},
\end{equation}

\begin{equation}
g_{1/2}^{-1}(y) \sim \exp\left[ -\frac{ \Gamma\left(1/2\right)^{2} }{ y^{2} } \right], \quad y \rightarrow \infty.
\end{equation}
By applying these expansions to the inequality of interest, we arrive at the following result:
\begin{equation}
\frac{\pi  \beta  \left(4 a_{1}^3 a_{12} \lambda_{2}^6 \left(a_{12}^2-a_{1} a_{2}\right)^2-\lambda_{1}^6 \left(a_{12}^4-a_{1}^2 a_{2}^2\right)^2\right)}{4 a_{1}^4 a_{12}^2 \lambda_{2}^{12}} >0\;,
\end{equation}
which leads to the following condition for the absence of a first-order phase transition:
\begin{equation}
    \left( \frac{\lambda_{1}}{\lambda_{2}} \right)^{6} < \frac{4 a_{1}^{3} a_{12} }{ \left( a_{12}^{2} + a_{1} a_{2} \right)^{2} }.
\end{equation}
To further analyze this result in the limit $D\rightarrow 0$, we rewrite the above condition as:
\begin{equation}
    \left( \frac{\lambda_{1}}{\lambda_{2}} \right)^{6} < \frac{4 a_{1}^{3} \sqrt{D + a_{1} a_{2}} }{ \left( D + 2a_{1} a_{2} \right)^{2} }\;,
\end{equation}
which for $D\to 0^-$ simplifies to:
\begin{equation}
    \left(\frac{\lambda_{1}}{\lambda_{2}}  \right)^{4} < \frac{a_{1}}{a_{2}}.
\end{equation}
The above condition describes the relation between the system parameters, which guarantees the absence of tricritical behaviour in type-1 particles. We can similarly derive conditions for type-2 particles. In the scenario, when the inequality saturates, a first-order phase transition remains absent. However, a tricritical point is shifted to zero temperatures, transforming it into a quantum tricritical point.

\subsection{Liquid-gas-type phase transition}

We now analyze the large-$D$ regime, where the imbalance leads to a qualitative change in the system phase diagram. As before, we fix the values of \(a_{12}\) and \(\beta\), and consider the imbalance parameters \(\delta = \frac{a_1}{a_2} \geq 1\) and \(\kappa = \frac{\lambda_2}{\lambda_1} \geq 1\).

As these imbalance parameters increase, a first-order phase transition emerges, separating the normal phase from the BEC$_{2}$. However, the character of the boundary between these phases changes. Rather than extending the boundary determined by Eq. (\ref{eq:BEC2boundary}), it now serves as a continuation of the liquid-gas-type phase transition - see Fig.\ref{fig:imbalance}.

\begin{figure}
    \centering
    \includegraphics[width=1\linewidth]{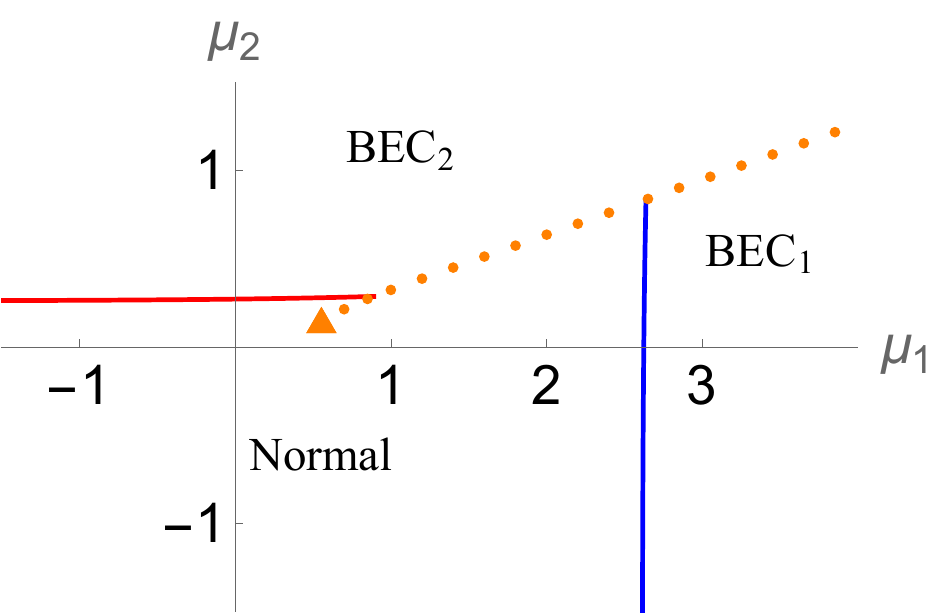}
    \caption{Projection of the phase diagram onto the $(\mu_{1}, \mu_{2})$ plane for sufficiently strong interspecies coupling $a_{12}$, where $D < 0$ and the interaction imbalance is $\frac{a_1}{a_2} = 10$. The red and blue lines denote continuous phase transitions from the normal phase to BEC$_2$ and BEC$_1$, respectively. The orange triangle mark the critical point. The orange dotted line represents a first-order transition: below the red line, it indicates a liquid-gas-type transition; above the red line and to the left of the blue line, it signifies the phase transition between the normal phase and BEC$_2$; to the right of the blue line, it denotes the phase transition between BEC$_2$ and BEC$_1$. The plot parameters are $\lambda_{1} = \lambda_{2}$, $\;\beta a_{12} \lambda^{-3} = 2$, $\;\beta a_{1} \lambda^{-3} = 1$, and $\;\beta a_{2} \lambda^{-3} = 0.1$.}
    \label{fig:imbalance}
\end{figure} 

When the imbalance becomes large enough, the point \(P_{2,c}\) shifts to the left of the intersection of the putative boundary lines. However, this placement alone no longer guarantees that $P_{2,c}$ has physical significance. This is due to its position to the right of the liquid-gas transition line. The situation can be better understood by envisioning the following scenario. In the absence of a liquid-gas-type transition, a first-order phase transition between the normal phase and a BEC$_{i}$ phase would require the presence of a tricritical point. Therefore, the relative positioning of $P_{i,c}$ to the putative boundary lines serves as a sufficient condition for this occurrence. 

Under the chosen imbalance conditions, the liquid-gas transition line intersects the phase boundary of BEC\(_{2}\). Thus, as we consider the BEC\(_{2}\) boundary, with \(P_{2,c}\) located to the right of the liquid-gas transition line, we observe that starting from negative values of \(\mu_{2}\) and moving along the boundary, there is only one solution at \(n_{2,c}\) until we reach the liquid-gas transition line.

At the intersection with the liquid-gas line, a degeneracy arises, yielding two possible solutions: one corresponds to the normal phase, and the other to BEC\(_{2}\). However, to the right of this intersection, the global minimum is associated with the normal phase. Thus, although a solution corresponding to BEC$_{2}$ exists at $P_{2,c}$, it is dominated by the normal-phase solution.

If the imbalance is increased even further, $P_{2,c}$ regains physical relevance by becoming a tricritical point. In this regime, two distinct first-order phase transitions emerge between the normal phase and BEC$_{2}$ one continuing the liquid-gas-type transition, and the other extending the boundary defined by Equation (\ref{eq:BEC2boundary}). However, for large imbalances, the liquid-gas-type transition does not occur.

Our key findings on the effects of interaction and mass imbalance can be summarized as follows: (i) We have demonstrated that, within a specific range of parameters, the first-order phase transition for one of the components can be suppressed entirely, regardless of the values of \(D\) or \(T\). (ii) The introduction of mass imbalance leads to a new type of first-order phase transition. Instead of extending the conventional phase boundary defined by Eq. (\ref{eq:BEC2boundary}), the mass imbalance creates an alternative scenario where the first-order phase transition occurs along the extension of the liquid-gas transition line.

\section{Summary and outlook}
The finite-$T$ phase diagram of simple Bose mixtures has been studied experimentally, theoretically, and by means of numerical simulations for a long time. Despite this, the picture that emerged over the years is incomplete (see in particular Ref.~\cite{Jakubczyk_2024}) and scattered. 
The point of the present paper is to clarify the situation at the mean-field level, at which we obtained new exact information concerning the global structure of the phase diagram, the stable thermodynamic phases, the order of the phase transitions and the multicritical points occurring (or not) in the system - see the summarizing paragraphs of each of the paper sections. We also discussed parameter regimes, which were hardly addressed before but are of increasing experimental relevance (such as the case of attractive interspecies interactions and the limit of collapse or the setups involving large mass or interaction imbalance). While for three-dimensional Bose gases, it is expected that the generic structure of the phase diagram should not be very sensitive to fluctuation effects, it is not clear if specific features, such as the order of the phase transitions, are retained when one goes beyond mean-field. The situation is completely different in two dimensions, where mean-field theory is largely inaccurate. In particular, fluctuation effects expel condensates featuring long-range order from the phase diagram, such that the low-temperature phases may only be of the Kosterlitz-Thouless type. Many questions concerning the two-dimensional Bose mixtures are open and constitute an interesting direction for future studies, which must necessarily go beyond the mean-field paradigms.

\begin{acknowledgments}
We are grateful to Marek Napi\'orkowski, F\'elix Rose, and Evgeny Sherman for their remarks on the initial version of the manuscript and discussions on related topics. We acknowledge support from the Polish National Science Center via grant 2021/43/B/ST3/01223. 
\end{acknowledgments}

\bibliography{bibliography.bib}
\bibliographystyle{apsrev4-1}

\end{document}